\documentstyle[12pt]{article}
\newcommand{\be}{\begin{equation}}
\newcommand{\bea}{\begin{eqnarray}}
\newcommand{\eea}{\end{eqnarray}}
\newcommand{\ba}{\begin{array}}
\newcommand{\ea}{\end{array}}
\newcommand{\ee}{\end{equation}}

\expandafter\ifx\csname mathbbm\endcsname\relax

\else

\fi
\def\l{\label}

\begin{document}
\begin{titlepage}
\hfill
\vbox{
    \halign{#\hfil         \cr
           IPM/P-2003/006 \cr
           hep-th/0302005  \cr
           } 
      }  
\vspace*{20mm}
\begin{center}
{\Large {\bf Semiclassical String Solutions on deformed NS5-brane 
Backgrounds and New Plane wave}\\ }

\vspace*{15mm}
\vspace*{1mm}
{Mohsen Alishahiha$^a$ \footnote{Alishah@theory.ipm.ac.ir}
and Amir E. Mosaffa$^{a,b}$}
\footnote{Mosaffa@theory.ipm.ac.ir} \\
\vspace*{1cm}

{\it$^a$ Institute for Studies in Theoretical Physics
and Mathematics (IPM)\\
P.O. Box 19395-5531, Tehran, Iran \\ \vspace{3mm}
$^b$ Department of Physics, Sharif University of Technology\\
P.O. Box 11365-9161, Tehran, Iran}\\

\vspace*{1cm}
\end{center}

\begin{abstract}
We study different Penrose limits of supergravity solution of NS5-brane in the
presence of RR field. Although in the case of NS5-brane we get a 
4-dimensional plane wave, in the case with RR field we will get two
different plane waves; a 4-dimensional and a 3-dimensional one. These
plane wave solutions are the backgrounds that a particular string solution 
feels at
one loop approximation. Using the one loop correction one can identify a 
particular subsector of LST/deformed LST which is dual to type II string
theories on these plane waves.

\end{abstract}

\end{titlepage}

\section{Introduction}

Little string theory (LST) is one of the most important examples of non-local
theories. This theory arises on the worldvolume of 
NS5-branes in a decoupling limit where $g_s \rightarrow 0$ with 
fixed $\alpha'$ \cite{{Berkooz:1997cq},{Seiberg:1997zk}}. LST's share many 
properties with usual string theory, such as T-duality and Hagedorn 
behavior of density of states, but are nevertheless non-gravitational 
theories. To study the theory one could use the  
holographic dual of LST, given in terms of string 
propagation in the linear dilaton background \cite{Aharony:1998ub}
along the AdS/CFT correspondence \cite{Maldacena:1997re}.
 
Recently there has been an improvement in understanding
this correspondence beyond the supergravity limit. 
In fact it has been conjectured \cite{Berenstein:2002jq}
that string theory on 
the maximally supersymmetric ten-dimensional plane wave has a 
description in terms of a certain subsector of the
large $N$ four-dimensional ${\cal N}=4\; SU(N)$ supersymmetric 
gauge theory at weak coupling. More precisely this subsector is 
parametrized by states with conformal weight $\Delta$ carrying 
$J$ units of charge under the $U(1)$ subgroup of the $SU(4)_{R}$ 
R-symmetry of the gauge theory, such that both $\Delta$ and $J$ 
are parametrically large in the large `t Hooft coupling while 
their difference, $\Delta-J$ is finite. Then it has been possible 
to work out the perturbative string spectrum from gauge theory 
side. The idea of \cite{Berenstein:2002jq} is based on the 
observation that the plane wave background with RR 4-form in type IIB 
string theory is maximally supersymmetric \cite{Blau:2001ne}, solvable
\cite{{Metsaev:2001bj},{Metsaev:2002re}} and can be understood as a 
certain limit (Penrose limit) of $AdS_5\times S^5$ geometry
\cite{{Blau:2002dy},{Berenstein:2002jq}}. 

On the other hand the authors of \cite{Gubser:2002tv} 
identified certain classical solutions representing highly 
excited string states carrying large angular momentum in the 
$AdS_5$ part of the metric with gauge theory operators with high 
spin and conformal dimension which is identified  
with the classical energy of the solution in the global
$AdS$ coordinates.  

In this context the BMN  operators \cite{Berenstein:2002jq} can 
also be identified with classical solutions of string in $AdS_5$ with 
angular momentum in $S^5$ space. Small fluctuations around this classical
solution in second order of fluctuation will lead to the corresponding
type IIB maximally supersymmetric plane wave 
\cite{{Gubser:2002tv},{Frolov:2002av}}. Then the one loop approximation around
this classical solution can be used to study the anomalous dimension of the 
BMN operators. Higher loop approximation can also be used to study the
higher order corrections to the anomalous dimension of BMN operators.
In fact from string theory point of view there would be corrections 
controlled by an expansion parameter given by ${1\over R^2}$, 
where $R^2$ is the radius of the AdS space. From the gauge theory point of 
view this leads to the ${1\over J}$ perturbative expansion to the anomalous 
dimensions of the BMN operators.

The idea of \cite{Gubser:2002tv} has been generalized for other 
string/M theory backgrounds in several papers including
\cite{{Russo:2002sr},{Pons:2003ci}}.

Physically the new scaling limit defined in 
\cite{Berenstein:2002jq} opens up new insight about the
gauge theory/string theory correspondence. Therefore it would be
natural to see whether this strategy can be used for other theories.
In fact this is the aim of this paper to explore this idea for LST
theory as well as its noncommutative/dipole deformations. We note,
however, that the Penrose limit of NS5-brane and its deformation has 
partially been studied in the literature, for example see
\cite{{Hubeny:2002vf},{Oz:2002ku},{Alishahiha:2002zu}}.

In this paper we will take the point of view of \cite{Gubser:2002tv}
to further study this theory. In section 2 we shall consider a  
classical string solution in the background generated by NS5-brane.
Using the quantum correction around this classical solution we have been 
able to identify a subsector of LST theory which is dual to the string theory
on a 4-dimensional plane wave. This subsector is parameterized by energy and 
angular momentum whose difference receives controllable quantum
corrections. 
In section 3 classical string solution in NS5-brane in the presence of nonzero 
RR field with all indices along the NS5-brane is studied. In this case
we find two different classes of classical solutions which lead to two
different plane waves. We will then identify the corresponding subsectors of
noncommutative LST theory which are dual to these plane waves. The one
loop approximation around these classical solutions has also been studied.
In section 4 the same analysis has been considered for dipole
deformation of LST. The last section is devoted to conclusions.

\section{ Plane wave from NS5-brane}

In this section we shall review the Penrose limit of the NS5-brane 
supergravity solution in type II string theories \cite{{Gomis:2002km},
{Kiritsis:2002kz},{Hubeny:2002vf}}. We will show that the corresponding 
plane wave is the background that a particular classical string solution 
feels in one loop approximation. This can be used to identify a subsector
of LST which is dual to string theory on the obtained plane wave. 

The supergravity solution on $N$ NS5-branes is given by
\bea
ds^2&=&-dt^2+d{\vec x}^2+f(dr^2+r^2d\Omega_3^2)\;,\cr &&\cr
e^{2\phi}&=&g_s^2f\;,\;\;\;\;\;dB=2Nl_s^2\epsilon_3\;,\;\;\;\;\;
f=1+{Nl_s^2\over r^2}\;,
\label{NS5}
\eea
where $ d\vec{x}$ parameterizes a 5-dimensional flat space and $\epsilon_3$ 
is the volume of $d\Omega_3^2$. In the decoupling limit
where $g_s\rightarrow 0$ and $l_s={\rm fixed}$\footnote{
Hereafter for simplicity, we set $l_s=1$.}
, setting $r=g_s u$, 
the supergravity solution (\ref{NS5}) reads
\bea
ds^2&=&-dt^2+d{\vec x}^2+{N\over u^2}(du^2+u^2d\Omega_3^2)\;,\cr 
&&\cr
e^{2\phi}&=&{N\over u^2}\;,\;\;\;\;\;dB=2N\epsilon_3\;,
\label{LST}
\eea
which provides the gravity dual description of LST.

To study the Penrose limit of the above supergravity solution 
we will first rescale 
$t\rightarrow \sqrt{N}t$ and consider a null geodesic
along $t$ and $\psi$ directions\footnote{Here we parameterize the 
3-sphere in (\ref{LST}) as
$d\Omega_3^{2}=d\eta_1^{2}+\cos^2{\eta_1}\;d\eta_2^{2}
+\sin^2{\eta_1}\;d\eta_3^{2}$.}  and at a fixed point with respect 
to other coordinates. In fact we consider a particle
moving along $\eta_2$ direction and sitting 
at $u=u_0$ and $\eta_1=0$. The geometry close to this trajectory
can be obtained by the following rescaling
\be
t={1\over 2}(x^++{x^-\over N}),\;\;\;\eta_2={1\over 2}(x^+-{x^-\over N}),
\;\;\;\;\eta_1={z\over \sqrt{N}},\;\;\;u=\sqrt{N}\;
e^{-\rho_0+\rho/ \sqrt{N}}\;.
\ee
In large $N$ limit keeping $x^+, x^-$ and $z$ fixed the solution(\ref{LST})
reads
\be
ds^2=-dx^+dx^--{1\over 4}z^2(dx^+)^2+d{\vec z}^2+d{\vec y}^2\;,
\;\;\;\;B_{+z_1}=z_2\;.
\l{PPLST}
\ee
with a constant dilaton $\phi=\rho_0$, and ${\vec y}=(\rho,{\vec x})$. 
This provides an exact string background to all orders in worldsheet 
perturbation theory \cite{{Amati:1988sa},{Horowitz:bv},{Nappi:1993ie}}.

Physically what we have considered is a physical system around the following 
classical string configuration 
\be
t=k\tau,\;\;\;\;\;\rho=\rho(\sigma),\;\;\;\;\;\eta_2=\omega \tau\;.
\ee
and all other coordinates are set to zero. This classical solution represents 
a string stretched along the radial 
coordinate and rotating along $\eta_2$ direction with the angular 
velocity $\omega$.
From the LST point of view this corresponds to a subsector of theory
parameterized by its energy\footnote{
If we want to work with the original coordinate as in (\ref{LST})
the energy $E$ needs to be scaled as $\sqrt{N}E$.}, $E$
and quantum number $J$ such that in large
$N$ limit both of them grow as $N$ while $E-J$ is finite.
Using the classical action of string  one can find a relation between
energy, $E$, and the angular momentum along the $\eta_2$ direction, $J$.
Doing so 
in the point like string limit one finds
\be
E^2={N\rho_0^2\over 4\pi^2}+J^2\;.
\l{EJ}
\ee
The background this state feels is the plane wave (\ref{PPLST}). 
To see this let us consider
small quantum fluctuations around this classical solution as following
\be
t=k\tau+{\tilde{t}\over \sqrt{N}}\;,\;\;\rho=\rho_0+{\tilde{\rho}\over 
\sqrt{N}}\;,\;\;\eta_2=
k\tau+{\tilde{\eta_2}\over \sqrt{N}}
\;,\;\;x_i=\tilde{x_i}\;,\;\;\eta_1={\tilde{z}_1\over \sqrt{N}}\;,\;\;
\eta_2=\tilde{z}_2
\ee 
Expanding the string action around this classical solution to
the second order in fluctuations, one finds: 
\bea 
I_{(2)}&=&
-{1\over4\pi}\int d^2\xi[-\partial_{a}\tilde{t}\partial^{a}\tilde{t} +
\partial_{a}\tilde{\eta_2}\partial^{a}\tilde{\eta_2}+
\partial_{a}\tilde{\vec{y}}\partial^{a}\tilde{\vec{y}}+
\partial_{a}\tilde{z}_{i}\partial^{a}\tilde{z}_{i}+k^2
\tilde{z}^{2}\cr
&&\;\;\;\;\;\;\;\;\;\;\;\;\;\;\;\;\;\;+
4k\tilde{z_2}\partial_{\sigma}\tilde{z_1}]\;.
\l{PPACTLST}
\eea
This action should be compared with the string action in the 
plane-wave background (\ref{PPLST}). To do this let us define
the lightcone coordinates $x^{\pm}=\eta_2\pm t$. In these coordinates
the expansion must be done around the following classical solution
\be
x^+=p^+\tau\;,\;\;\;x^-=\vec{y}=z_i=0\;,\;\;\;\rho=\rho_0\;,\;\;\;p^+=2k\;.
\ee
One can see that in the quantum gauge, $\tilde{x}^{+}=0$, the
action (\ref{PPACTLST}) reads
\bea
I_{(pw)}&=& -{1\over4\pi}\int
d^2\xi [\partial_{a}x^{+}\partial^{a}x^{-}-
{1\over4}z_i^{2}\partial_{a}x^+\partial^{a}x^+ +
\partial_{a}\vec{y}\partial^{a}\vec{y}+
\partial_{a}z_i\partial^{a}z_i\cr &&\;\;\;\;\;\;\;\;\;\;\;\;\;\;\;\;\;
+2B_{+z_1}\partial_{\tau}x^{+}\partial_{\sigma}z_1]\;,
\eea 
which is the string action on the plane wave 
background (\ref{PPLST}) in the lightcone gauge.

The energy, $E$ and the angular momentum, $J$ of the point like string up to 
the second order will read: 
\bea 
&E&={1\over2\pi}\int d\sigma
(k+\partial_{\tau}\tilde{t})\;,\;\;\;\;\;\;\;\;\;\;
\cr&J&={1\over2\pi}\int d\sigma
(k+\partial_{\tau}\tilde{\eta_2}-k\tilde{z}^2-\tilde{z_2}
\partial_{\sigma}\tilde{z_1})\;,
\eea 
therefore we get 
\be 
E-J={1\over4\pi}\int
d\sigma(-2\partial_{\tau}\tilde{x}^{-}+2k\tilde{z}^2+
2\tilde{z}_2\partial_{\sigma}\tilde{z_1})\;.
\ee
By making use of the Virasoro constraint, 
\bea
2k\partial_{\tau}\tilde{x}^{-}&-&\partial_{\tau}\tilde{t}
\partial_{\tau}\tilde{t}-
\partial_{\sigma}\tilde{t}\partial_{\sigma}\tilde{t}+
\partial_{\tau}\tilde{\vec{y}}\partial_{\tau}\tilde{\vec{y}}+
\partial_{\sigma}\tilde{\vec{y}}\partial_{\sigma}\tilde{\vec{y}}+
\partial_{\tau}\tilde{z_i}\partial_{\tau}\tilde{z_i}+
\partial_{\sigma}\tilde{z_i}\partial_{\sigma}\tilde{z_i}\cr&+&
\partial_{\tau}\tilde{\eta_2}\partial_{\tau}\tilde{\eta_2}+
\partial_{\sigma}\tilde{\eta_2}\partial_{\sigma}\tilde{\eta_2})-
k^2\tilde{z}^2=0,
\eea
one can solve $\partial_{\tau}\tilde{x}^{-}$ in terms of the second 
order fluctuations, and thus we find
\bea 
E-J={1\over4\pi k} \int
d\sigma&[&\partial_{\tau}\tilde{x}^{+}\partial_{\tau}\tilde{x}^{-}+
\partial_{\sigma}\tilde{x}^{+}\partial_{\sigma}\tilde{x}^{-}+
\partial_{\tau}\tilde{\vec{y}}\partial_{\tau}\tilde{\vec{y}}+
\partial_{\sigma}\tilde{\vec{y}}\partial_{\sigma}\tilde{\vec{y}}\cr&+&
\partial_{\tau}\tilde{z_i}\partial_{\tau}\tilde{z_i}+
\partial_{\sigma}\tilde{z_i}\partial_{\sigma}\tilde{z_i}+
k^2\tilde{z}^2+4k\tilde{z_2}\partial_{\sigma}\tilde{z_1}]\;. 
\l{EJLST}
\eea
The expression in the right hand side is the 2-dimensional transverse 
Hamiltonian of the worldsheet. 
By making use of this expression one can find the quantum correction to 
the classical relation between $E$ and $J$. In fact one finds
\bea
E-J&=&\sum_{n}\left({N|n|\over J}N_n^{(6)}+\omega^+_nN_n^++\omega^-_nN_n^-
\right)\cr
&&\cr \omega^{\pm}_n&=&|1\pm {Nn\over J}|,
\l{aaaa}
\eea
where $N_n^{(6)},N_n^+$ and $N_n^+$ are occupation numbers along the
6-dimensional
flat space and directions $z_1$ and $z_2$ when they are written in normal
modes. 

Therefore one
might conclude that type IIA string theory on the plane wave background
(\ref{PPLST})
is dual to a subsector of LST theory which is parameterized by energy, $E$ and
angular momentum, $J$ such that both of them grow like $N$ in large $N$ limit
while $E-J$ is finite and given by (\ref{aaaa}).

\section{Plane wave from noncommutative deformation of NS5-brane}

In this section we shall consider the Penrose limit of noncommutative
deformation 
of LST theory. This will be done by making use of the supergravity solution of 
NS5-brane in the presence of nonzero RR field. We will first consider the case
where all indices of RR fields are along the NS5-brane worldvolume. In fact 
these are decoupled theories (ODp) on the worldvolume of 
type II NS5-branes in the presence of nonzero $p$-form RR filed
\cite{{OM},{HARM1}}.\footnote{For $p=1,2$ see also 
\cite{{Alishahiha:1999ci},{ALI1}}.}
The excitations of these theories include light open Dp-branes.
The gravity description of these theories have been studied in
\cite{AOR,mitra-roy}. We could also
consider a case where one leg of the RR fields is in the transverse
direction which
corresponds to the dipole deformation of LST. This case will be studied in the 
next section.
      
The gravity description of ODp theory is given by \cite{AOR}
\bea
ds^2&=&(1+a^2r^2)^{1/2}\left[-dt^2+\sum_{i=1}^{p}
dx_{i}^2+\frac{\sum_{j=p+1}^{5}d x_{j}^2}{1+a^2r^2}
+{ N\over r^2}(dr^2+r^2d\Omega^2_3)\right]\ ,\cr
&&\cr
A_{0\cdots p}&=&{1 \over {\tilde g}} a^2r^2\ \ ,
\;\;\;\;\;\;\;\;\;\;
\;\;\;\;\;\;
A_{(p+1)\cdots 5}={1\over {\tilde g}}\;\frac{a^2r^2}{1+a^2r^2}\ ,
\cr
&&\cr
e^{2\phi}&=&{\tilde g}^2 \frac{(1+a^2r^2)^{(p-1)/2}}{a^2r^2},
\;\;\;\;\;\;\;dB=2N\;\epsilon_3,\;\;\;\;a^2={l_{\rm eff}^2\over N}\;,
\label{rrrr}
\eea
where $l_{\rm eff}$ and ${\tilde g}$ are effective string tension and 
effective string coupling of the theory which are the parameters of 
the theory after taking the decoupling limit \cite{OM}. 
Here $\epsilon_3$ is the volume of $S^3$ part of the metric.

\subsection{4-dimensional plane wave}

The Penrose limit of background (\ref{rrrr}) has been studied in 
\cite{{Oz:2002ku},{Alishahiha:2002zu}} (see also 
\cite{{Bhattacharya:2002zf},{Bhattacharya:2002qx},{Fuji:2002vs}}).
The obtained plane wave in this case corresponds to the background
which a point like string moving in the 3-sphere feels. In fact
to get the corresponding plane wave we first rescale $t\rightarrow \sqrt{N}t$ 
and
then consider the following classical string configuration
\be
t=k\tau,\;\;\;\;\;\rho=\rho(\sigma),\;\;\;\;\;\eta_2=\omega \tau\;,
\l{CLSS2}
\ee
where $\rho=\ln(ar)$. The Nambu-Gotto action for this classical string
configuration reads
\be
I=-{1\over 2\pi}\sqrt{N(k^2-\omega^2)}\int d\rho \sqrt{1+e^{2\rho}}\;,
\ee
which in the point like string limit leads to the similar result as 
(\ref{EJ}). Actually the plane wave background seen by the small quantum
fluctuations 
around this classical solution can be obtained from a null geodesic
around $\rho=\rho_0$=constant \cite{Alishahiha:2002zu}. In this case
we perform the following coordinate transformations:
\be\ba {ll}
\rho =\rho_0 + ( 1 + e^{2\rho_0})^{-1/4}{r\over \sqrt{N}}, &
x_j=( 1 + e^{2\rho_0})^{1/4}w_j,\;\;(j=p+1,..,5),\cr
\eta_1 =( 1 + e^{2\rho_0})^{-1/4}{z \over \sqrt{N}}, &
x_i = ( 1 + e^{2\rho_0})^{-1/4}y_i\;\;(i=1,..,p),\cr
t ={1\over 2}( 1 + e^{2\rho_0})^{-1/4}(x^+ + {x^-\over N}),&
\eta_2 ={1\over 2} ( 1 + e^{2\rho_0})^{-1/4}(x^+ - {x^-\over N})\;,
\ea\ee
and $\eta_3$ is kept fixed. One then obtains in the Penrose limit 
($N\rightarrow \infty$) a plane wave background as following
\be
ds^2 = -dx^+ dx^- - {z^2\over 4} {dx^+}^2 + d{\vec z}^2+d{\vec y}^2\;,
\label{RNSPP}
\ee
where $d{\vec y}^2=dr^2+d{\vec w}^2+d{\vec y}^2$.
Here we have also rescaled the longitudinal coordinates by
$x^{\pm}\rightarrow x^{\pm} (1 + e^{2\rho_0})^{\pm{1/4}}$. 
In this limit, one gets also a nonvanishing $p+1$ form field and nonzero
B field:
\bea
dA_{+ry_1\cdots y_p} &=& {1\over \tilde{g}}\; e^{2\rho_0} 
(1+ e^{2\rho_0})^{-(p+1)/4}\;,\cr &&\cr
A_{(P+1)\cdots 5}&=&{1\over {\tilde g}}\;\frac{e^{2\rho_0}}{1+e^{2\rho_0}}
\;,\cr &&\cr
B_{+z_1}&=&(1+e^{2\rho_0})^{-1/2}z_2\;,
\label{RNSPPF}
\eea
and the constant dilaton after the Penrose limit is given by:
\be
e^{2\phi} = \tilde{g}^2 {(1 + e^{2 \rho_0})^{(p-1)/2}\over
e^{2\rho_0}}\;.
\ee

To compare this plane wave solution with that obtained from NS5-brane 
supergravity
background it is constructive to study the small quantum fluctuations around
the above classical solution. Consider the following small fluctuations
around the
classical solution (\ref{CLSS2})
\be 
t=k\tau+\tilde{t},\;\;\;\;\eta_2=k\tau+\tilde{\eta_2},\;\;\;
 \rho=\rho_0+\tilde{\rho},\;\;\;x_i=\tilde{y_i},
 \;\;\;x_j=\tilde{y_j},\;\;\;\eta_i=\tilde{z_i}\;,
\ee 
with $i=1,\cdots,p$ and $j=p+1,\cdots,5$.
For simplicity we first rescale the coordinates as follows: 
\bea 
&&\tilde{t} \rightarrow
{(1+e^{2\rho_0})^{-1/4}\over\sqrt{N}}\tilde{t},\;\;\;\;\tilde{x}_{i}\rightarrow
(1+e^{2\rho_0})^{-1/4}\tilde{y}_{i},\;\;\;\;\tilde{x}_{j}\rightarrow
(1+e^{2\rho_0})^{1/4}\tilde{y}_{j},\cr &&\cr
&&\tilde{\rho}\rightarrow
{(1+e^{2\rho_0})^{-1/4}\over\sqrt{N}}\tilde{\rho},\;\;\; \tilde{z}_{1}
\rightarrow  {(1+e^{2\rho_0})^{-1/4}\over\sqrt{N}}\tilde{z}_{1},\;\;\;
\tilde{\eta}_2\rightarrow
{(1+e^{2\rho_0})^{-1/4}\over\sqrt{N}}\tilde{\eta}_2\;, 
\eea
and $\tilde{z}_{2}$ is also rescaled by $(1+e^{2\rho_0})^{-1/4}$.
We can now proceed as in the previous
section to write the string action up to the second order in fluctuations.
The result is:
\bea
I_{(2)}=-{1\over4\pi}\int
d^2\xi\;[&-&\partial_{a}\tilde{t}\partial^{a}\tilde{t}+
\partial_{a}\tilde{\eta}_{2}\partial^{a}\tilde{\eta}_{2}+
\partial_{a}\tilde{\vec{y}}_{i}\partial^{a}\tilde{\vec{y}}_{i}+
\partial_{a}\tilde{z}_{i}\partial^{a}\tilde{z}_{i} \cr&+&
k^2 \tilde{z}^2+{4k\over(1+e^{2\rho_0})^{1/2}}\tilde{z_2}
\partial_{\sigma}\tilde{z_1}]\;,
\label{action1} 
\eea 
where $\vec{y}=(\rho,x_i,x_j)$. To compare this action with the 
string action on plane wave solution (\ref{RNSPP}) we define the 
lightcone coordinates as $x^{\pm}=\eta_2\pm t$. In these 
coordinates the small fluctuations must occur around the 
following classical solution
\be
x^{+}=p^{+}\tau\;,\;\;\;x^{-}=0\;,\;\;\;x_{i}=x_{j}=0\;,\;\;\;
\rho=\rho_{0}\;,\;\;\;z_{i}=0\;,\;\;\;p^{+}=2k\;.
\ee
Then in the quantum gauge $\tilde{x}^{+}=0$ the action
(\ref{action1}) reads
\bea 
I_{(pw)}=-{1\over4\pi}\int
d^2\xi&[&
\partial_{a}x^{+}\partial^{a}x^{-}-
{1\over4}z^{2}\partial_{a}x^{+}\partial^{a}x^{+}+
\partial_{a}y_{i}\partial^{a}y_{i}+\partial_{a}z_{i}\partial^{a}z_{i}\cr&+&
2B_{+z_1}\partial_{\tau}x^{+}\partial_{\sigma}z_1]\;, 
\eea
with $B_{+z_1}=(1+e^{2\rho_0})^{-1/2}z_2$. We recognize this action as 
the string action on the plane wave solution (\ref{RNSPP}).

As in the previous section one can write the Virasoro constraint and 
solve it for $\partial_{\tau}\tilde{x}^{-}$ in terms of the second order
fluctuations. This can be used to find an expression for 
the energy and angular momentum up to second order in fluctuations.
Using the obtained expressions for energy and angular momentum one finds  
\bea 
E-J=
{1\over4\pi k}\int
d\sigma&[&\partial_{\tau}\tilde{x}^{+}\partial_{\tau}\tilde{x}^{-}+
\partial_{\sigma}\tilde{x}^{+}\partial_{\sigma}\tilde{x}^{-}+
\partial_{\tau}\tilde{y}_{i}\partial_{\tau}\tilde{y}_{i}+
\partial_{\sigma}\tilde{y}_{i}\partial_{\sigma}\tilde{y}_{i}\cr&+&
\partial_{\tau}\tilde{z}_{i}\partial_{\tau}\tilde{z}_{i}+
\partial_{\sigma}\tilde{z}_{i}\partial_{\sigma}\tilde{z}_{i}+
k^2\tilde{z}^{2}+ {4k\over (1+e^{2\rho_0})^{1/2}}\tilde{z_2}
\partial_{\sigma}\tilde{z_1}]\;. 
\l{EJDLST1}
\eea 
Note that the right side in the above equation is the worldsheet
Hamiltonian of the transverse fluctuations. Therefore we get\footnote{
In writing this expression we have used the relation between string theory
parameters and worldvolume parameters as $P^+=(1+e^{2\rho_0})^{-1/2}
{2J\over N}$.}
\bea
E-J&=&\sum_{n}\left[\sqrt{1+e^{2\rho_0}}
{N|n|\over J}N_{n}^{(6)}+\omega^{+}_nN_n^++\omega^-_nN_n^-\right]\;,\cr &&\cr
\omega^{\pm}_n&=&\sqrt{1+(1+e^{2\rho_0}){N^2n^2\over J^2}
\pm2{Nn\over J}}\;,
\l{EJ2}
\eea
where $N_n^{(6)},N_n^+$ and $N_n^-$ are occupation numbers along the 
6-dimensional flat space and directions $z_1$ and $z_2$ when they are
written in normal modes, respectively.
Thus string theory on the plane wave background (\ref{RNSPP}) is dual
to a subsector of the noncommutative deformation of LST which is 
parameterized by energy $E$ and a quantum number $J$ such that in 
large $N$ limit both of them grow as $N$ while $E-J$ is finite
and given by (\ref{EJ2}).

\subsection{3-dimensional plane wave}

In comparison with the NS5-brane background the deformed NS5-brane 
background (\ref{rrrr}) has an extra parameter corresponding to the
RR fields. This parameter can also be thought of as the deformation
parameter $a$. The presence of RR field will also break the Lorentz
symmetry. Therefore it is natural to consider a boost along the
direction in which the RR field is defined. This can be thought of 
as an other Penrose limit of the background. Physically, what we are
considering is a fast moving particle along those directions where the 
RR field is defined. Then we can look at the theory close to the
trajectory of this particle which leads to a new plane wave background. 
In fact what we really have to do is to consider a classical string 
configuration which is stretched along the radial coordinate and moves 
in a direction where the RR-field is defined. 
In the point like limit this would be the physical system we are looking 
for.

Let us first obtain the corresponding plane wave and then study the
semiclassical string solution leading to this plane wave. 
To find the corresponding plane wave we consider the following 
rescaling
\be
t={1\over 2}\left({x^+\over a}+ax^-\right),\;\;\;\;
x_5={1\over 2}\left({x^+\over a}-ax^-\right).
\ee
In the limit of $a\rightarrow 0$ keeping $g_0={\tilde g}/a$ fixed the
supergravity
solution (\ref{rrrr}) for $p\neq 5$ reads
\bea 
ds^2&=& -dx^+dx^--{r^2\over 4}(dx^+)^2+{N\over r^2}dr^2+d{\vec
y}^2+Nd\Omega_3^2,
\;\;\;\;\;\;dB=N\epsilon_3,\cr &&\cr
A_{+1\cdots p}&=&{1\over 2g_0}\;r^2,\;\;\;\;\;
A_{(p+1)\cdots 4+}={1\over 2g_0}\;r^2,\;\;\;\;\;\;e^{2\phi}={g_0^2\over r^2}\;.
\l{LNS}
\eea
The case of $p=1$ has first been studied in \cite{AOR} in the context of 
lightlike noncommutative geometry. In fact the dual theory we get from 
this Penrose limit is the lightlike noncommutative deformation of LST. 
We note also that the obtained plane wave provides a string theory background 
in which the string theory can be exactly solved. Actually this consists of 
level $N$ $SU(2)$ WZW, a three dimensional Liouville plane wave plus 4 free 
fields theory. The Liouville plane wave background in string theory has 
recently 
been considered in \cite{{Maldacena:2002fy},{Russo:2002qj}} as a background 
in which the string theory can be exactly solved.

The classical string solution representing a string moving along $x_5$
direction 
and stretched in the radial coordinate is
\be
t=k\tau,\;\;\;\;\;r=r(\sigma),\;\;\;\;\;x_5=p\tau\;.
\ee
For point like strings and in the limit where $\;a\rightarrow
0\;$ one find a relation between energy $E$ and momentum $P$ which 
is similar to (\ref{EJ}). We can also study the small quantum fluctuations 
around this classical solution. Basically the procedure is very similar to 
what we have considered in the previous cases. The only point in this case 
is that the WZW part of the action remains unchanged in the procedure 
of the expansion. In fact the bosonic part of the action takes the
following form
\be
I_{\rm total}=I(t,r,x_i,x_j)+ I_{\rm WZW}
\ee
up to an ``$a$'' dependent coefficients in WZW part which is one
in $a\rightarrow 0$ limit. The first term is also given by
\be
I=-{1\over 4\pi}\int d^2\xi\bigg{[}(1+a^2r^2)^{1/2}(-\partial_\alpha
t\partial^\alpha t
+\partial_\alpha x_i\partial^\alpha x_i+{\partial_\alpha x_j\partial^\alpha x_j
\over 1+a^2r^2}
+{N\over r^2}\partial_\alpha r\partial^\alpha r)\bigg{]} \;,
\ee
which up to second order in the fluctuations around the above classical
solution, $t=k\tau+{\tilde t}, x_5=k\tau+{\tilde x_5}, r={\tilde r}$, reads
\be
I_{(2)}=\int d^2\xi\bigg{[}-\partial_\alpha {\tilde t} \partial^\alpha 
{\tilde t}
+\partial_\alpha {\tilde x}_5\partial^\alpha {\tilde x}_5+{N\over {\tilde r}^2}
\partial_\alpha {\tilde r}\partial^\alpha {\tilde r}+
\partial_\alpha {\tilde y}_i\partial^\alpha {\tilde y}_i
+m^2{\tilde r}^2\bigg{]}\;,
\ee
where $y_i,\;i=1,\cdots, 4$ represents the four transverse directions to the
plane wave. Note also that to get the above action we have taken into
account the limit of $a\rightarrow 0$ while kept $m:=ka$ fixed.
It is easy to see that this action plus the WZW part is the
bosonic part action of string in the plane wave background (\ref{LNS}).
One could proceed to compute $E$ and $J$ and thereby to find
the quantum correction to $E-J$. This would of course get
corrections from different parts of the action; contributions from
4 free fields theory, 3-dimensional Liouville plane wave and $SU(2)$
WZW. It would be interesting to find each contribution explicitly.

\section{Plane wave from dipole deformation of NS5-brane}

The dipole deformation of LST can be described by the
supergravity solution of type IIB NS5-branes in the 
presence of an RR 2-form potential with one leg along the brane and the
other along the transverse directions\footnote{For definition 
of dipole field theory and its relevance to string theory
see, for example, \cite{Bergman:2000cw}-\cite{Bergman:2001rw}.}. 
We can also make a series of T-duality transformations
to produce a new supergravity solution. This supergravity
solution describes type II NS5-branes 
in the presence of RR $(6-p)$-form, for $p=0\dots 4$, with
one leg along the transverse directions and $(5-p)$ legs along
the NS5-branes worldvolume. The corresponding supergravity solution
in the decoupling limit is given by \cite{Alishahiha:2002ex}
\bea
ds^2&=&(1+r^2L^2)^{1/2}\bigg{[}dt^2-\sum_{i=1}^{p}dx_i^2-
\frac{\sum_{j=p+1}^5dx_j^2}
{1+r^2L^2} \cr &&\cr 
&-&{N\over r^2}\bigg{(}dr^2+r^2d\Omega_3-\frac{r^4L^2}
{1+r^2L^2}(a_1d\theta_1+a_2d\theta_2  
+  a_3d\theta_3)^2\bigg{)}\bigg{]}\;,
\cr &&\cr
e^{2\phi}&=&\frac{N}{ r^2}(1+r^2L^2)^{(p-2)/2}\;,\cr &&\cr
\sum_{a=6}^{9}A_{(p+1)\cdots 5\theta_a}d\theta_a&=&\frac{r^2L}
{1+r^2L^2}(a_1d\theta_1+a_2d\theta_2+a_3d\theta_3)\;,
\label{SUGNS5}
\eea
where $\theta_i$'s are angular coordinates parameterizing the sphere 
$S^{3}$ transverse to the NS5-branes\footnote{Note that here we used a 
parameterization such that in terms of $\theta_i$'s the metric of
3-sphere is given by $d\Omega_3^2=d\theta_1^2+\cos^2\theta_1\;
d\theta_2^2+\cos^2\theta_1\cos^2\theta_2\;d\theta_3^2$.}, and
\be
a_1= \cos\theta_2,\;\;\;\;\;
a_2= -\sin \theta_1 \cos \theta_1 \sin \theta_2,\;\;\;\;\;
a_3=\sin^2 \theta_1 \sin^2 \theta_2\;.
\label{ASM5}
\ee
Here $L$ is the effective dipole moment. There is also a two form 
$B$ field representing the charge of the NS5-branes which is given 
by $dB=2N\epsilon_3$. Note that the above solution is 
maximally supersymmetric which means that the solution preserves 8 supercharges.

\subsection{4-dimensional plane wave}
Let us first study the system close to the trajectory of a point like
string moving in a direction of 3-sphere transverse to the NS5-brane.
The background this state feels would be a plane wave solution. To find this
plane wave 
we first rescale $t\rightarrow \sqrt{N}t$ and define a new
radial coordinate $\rho$ as $Lr=e^{\rho}$. In these coordinates the
supergravity solution (\ref{SUGNS5}) reads
\bea
ds^2&=&(1+e^{2\rho})^{1/2}\bigg{[}-Ndt^2+\sum_{i=1}^{p}dx_i^2+
\frac{\sum_{j=p+1}^5dx_j^2}
{1+e^{2\rho}} \cr &&\cr 
&+&Nd\rho^2+Nd\Omega_3-\frac{Ne^{2\rho}}
{1+e^{2\rho}}(a_1d\theta_1+a_2d\theta_2  
+  a_3d\theta_3)^2\bigg{]}\;,
\cr &&\cr
e^{2\phi}&=&g_s^2\frac{(1+e^{2\rho})^{(p-2)/2}}{e^{2\rho}}
\;,\cr &&\cr
\sum_{a=1}^{3}A_{(p+1)\cdots 5\theta_a}d\theta_a&=&\frac{e^{2\rho}}
{L(1+e^{2\rho})}(a_1d\theta_1+a_2d\theta_2+a_3d\theta_3)\;.
\l{bbbb}
\eea
Now we will consider a null geodesic in $(t,\theta_3)$ directions around 
$\rho=\rho_0$=constant. In this case we perform the following coordinate
transformations 
\be\ba {ll}
\rho =\rho_0 + ( 1 + e^{2\rho_0})^{-1/4}{r\over \sqrt{N}}, &
x_j=( 1 + e^{2\rho_0})^{1/4}w_j,\;\;(j=p+1,..,5),\cr
\theta_{\alpha} =( 1 + e^{2\rho_0})^{-1/4}{z_{\alpha} \over \sqrt{N}}, &
x_i = ( 1 + e^{2\rho_0})^{-1/4}y_i\;\;(i=1,..,p),\cr
t ={1\over 2}( 1 + e^{2\rho_0})^{-1/4}(x^+ + {x^-\over N}),&
\theta_3 = {1\over 2}( 1 + e^{2\rho_0})^{-1/4}(x^+ - {x^-\over N})\;,
\ea\ee
with $\alpha=1,2$. In  the limit of $N\rightarrow \infty$ we 
find the following plane wave from the supergravity solution (\ref{bbbb})
\bea
ds^2&=&-dx^+dx^--{1\over 4}\left((1+e^{2\rho_0})z_1^2+z_2^2\right) 
(dx^+)^2+d{\vec z}^2 +d{\vec y}^2\;,\cr
e^{2\phi}&=&g_s^2\frac{(1+e^{2\rho_0})^{(p-2)/2}}{e^{2\rho_0}},\;\;\;\;\;
B_{+z_1}=z_2 \;,
\l{DIPP1}
\eea
where $d{\vec y}^2=dr^2+dy_i^2+dw_j^2$. This 4-dimensional plane wave is very
similar to that in LST case. Of course unlike the LST case, in this case 
$z_1$ and $z_2$ have different masses which is the effect of dipole deformation.

The corresponding classical string solution is  
\be
t=k\tau\;\;\,\;\;\;\rho=\rho(\sigma)\;\;,\;\;\;\theta_3=\omega\tau\;,
\ee 
which in the point like limit leads to the above plane wave at second order 
in fluctuations. To see this we consider the small fluctuations around this 
classical solution. In the one loop approximation one gets
\bea 
I_{2}&=&-{1\over 4\pi}\int d^2\xi
\bigg{[}\partial_{a}\tilde{x}^{+}\partial^{a}\tilde{x}^{-}+
\partial_{a}\tilde{\vec{y}}\partial^{a}\tilde{\vec{y}}+
\partial_{a}\tilde{\theta}_{i}\partial^{a}\tilde{\theta}_{i}+
k^2\left((1+e^{2\rho_0})\tilde{\theta}_{1}^2+\tilde{\theta}_{2}^2\right)\cr&&
\;\;\;\;\;\;\;\;\;\;\;\;\;\;\;\;\;\;\;
2k\tilde{\theta}_{2}\partial_{\sigma}\tilde{\theta}_{1}\bigg{]} \;,
\eea
which is equivalent to the bosonic part of the string action on the
plane wave background (\ref{DIPP1}).
As in the previous cases one can also write the expressions for 
energy and angular momentum in terms of the transverse modes. Using the
Virasoro constraint in the one loop approximation one finds: 
\bea
E-J&=&{1\over 2\pi k}\int d\sigma\bigg{[}
\partial_{\tau}\tilde{x}^{+}\partial_{\tau}\tilde{x}^{-}+
\partial_{\sigma}\tilde{x}^{+}\partial_{\sigma}\tilde{x}^{-}+
\partial_{\tau}\tilde{\vec{y}}\partial_{\tau}\tilde{\vec{y}}+
\partial_{\sigma}\tilde{\vec{y}}\partial_{\sigma}\tilde{\vec{y}}+
\partial_{\tau}\tilde{\theta}_{i}\partial_{\tau}\tilde{\theta}_{i}\cr &&
\;\;\;\;\;\;\;\;\;\;\;\;\;\;\;\;\;\;+
\partial_{\sigma}\tilde{\theta}_{i}\partial_{\sigma}\tilde{\theta}_{i}+
k^2[(1+e^{2\rho_0})\tilde{\theta}_{1}^{2}+\tilde{\theta}_{2}^{2}]+
4k\tilde{\theta}_{2}\partial_{\sigma}\tilde{\theta}_{1}\bigg{]}\;.
\eea
We note that the right hand side is the Hamiltonian of the transverse
fluctuations. Therefore we find
\be
E-J=\sum_{n}\bigg{[}\sqrt{1+e^{2\rho_0}}{N|n|\over J}N_n^{(6)}
+\omega^+_nN_n^++\omega^-_nN_n^-\bigg{]}\;,
\l{EJDipole}
\ee
where $N_n^{(6)}$ is the occupation number of the 6 transverse directions and
$N_{n}^{\pm}$ is occupation number along $z_1$ and $z_2$ when they are written 
in normal mode representation. The $\omega^{\pm}$ is given by
\be
\omega^{\pm}_n=\left(1+{e^{2\rho_0}\over 2}+(1+e^{2\rho_0}){N^2n^2\over J^2}
\pm\sqrt{(1+e^{2\rho_0}){4N^2n^2\over J^2}+{e^{4\rho_0}\over
4}}\;\right)^{1/2}\;.
\ee 
Thus one may conclude that string theory on the plane wave background
(\ref{DIPP1}) 
is dual to a subsector of dipole deformation of LST parameterized by
energy, $E$ 
and angular momentum, $J$ which at large $N$ limit both of them grow as
$N$ while
$E-J$ is finite and given by (\ref{EJDipole}).

\subsection{3-dimensional plane wave}
Let us now study a point like string moving in $x_5$ direction and look at
the physics close to the trajectory of this point like string. 
To study the system near this trajectory we perform the following 
rescaling
\be
t={1\over 2}\left({x^+\over L}+Lx^-\right),\;\;\;\;
x_5={1\over 2}\left({x^+\over L}-Lx^-\right).
\ee
In the limit of of $L\rightarrow 0$ the supergravity solution (\ref{SUGNS5})
reads
\bea
ds^2&=&-dx^+dx^--{r^2\over 4}(dx^+)^2
+N{dr^2\over r^2}+Nd\Omega_3^2+\sum_{i=1}^4dx_i^2\;,\cr &&\cr
e^{2\phi}&=&{N\over r^2},\;\;\;\;\;
\sum_aA_{(p+1)\cdots 4 \theta_a +}d\theta_a
=-{r^2\over 2}(a_1d\theta_1+a_2d\theta_2+a_3d\theta_3)\;,
\label{LIO}
\eea
and the B field remains unchanged.

This plane wave solution has been studied in \cite{Alishahiha:2003ru}
in the context of lightlike dipole deformation of LST theory
(see also \cite{Ganor:2002ju}). Note that
the metric, dilaton and B field in  the plane wave solution (\ref{LIO}) are
the same as in the plane wave solution coming from lightlike
noncommutative deformation of LST theory (\ref{LNS}) and the only difference
is in the RR fields. This means that as far as the bosonic part of the 
string theory on these backgrounds is concerned both of them give the same 
result. But of course the fermionic parts will be different. It is
also worth noting that there is a nonzero RR field with one leg along
the the direction where the WZW part of the theory is defined, {\it i.e.} the
3-sphere. This could also deform the $SU(2)$ WZW model as well. Therefore
the three different parts of the theory contributing to the correction of $E-J$
are as following: 4 free fields theory, 3-dimensional Liouville plane wave and
deformed $SU(2)$ WZW. Since the RR field has only one leg along the 3-sphere 
one might suspect that the corresponding deformation is dipole deformation. 
It would be interesting to study this model in more detail.

\section{Conclusion}

In this paper we have studied different Penrose limits of type II NS5-brane 
solution in the presence of different RR fields. We have seen that although
for NS5-brane we get only one plane wave, for the case with RR field two 
different plane waves can be obtained. Interestingly enough all of these plane 
waves
lead to exactly solvable string theory backgrounds.

We have also considered the semiclassical string configuration on the
backgrounds generated by NS5-brane as well as NS5-brane in the presence
of different nonzero RR fields. These string solutions give at one loop 
approximation the obtained plane wave from the corresponding geometry. 
This has been used 
to identified a subsector of LST/deformed LST which is dual to the string
theory in the corresponding plane wave background. 

For all cases there is a Penrose limit leading to a 4-dimensional plane wave. 
In this case the subsector of LST, noncommutative LST and dipole LST is 
parameterized by energy and angular momentum. In all cases both energy and
angular momentum grow as $N$ in large $N$ limit while their differences
remain finite. In fact we have written a closed form expression for $E-J$ in 
all cases. Actually, we have found that the subsector of LST theory is 
a set of operators carrying angular momentum $J$ of $U(1)$ subgroup
of the global symmetry $SO(4)$ with energy $E$ such that
\bea
E-J&=&\sum_{n}\left[\sqrt{1+e^{2\rho_0}}
{N|n|\over J}N_{n}^{(6)}+\omega^{+}_nN_n^++\omega^-_nN_n^-\right]\;,\cr &&\cr
\omega^{\pm}_n&=&\sqrt{1+(1+e^{2\rho_0}){N^2n^2\over J^2}
\pm2{Nn\over J}}\;.
\eea 

To compare the subsector of deformed LST with the subsector of LST  
we observe that the factor $1+e^{2\rho_0}$ in (\ref{EJ2}) and (\ref{EJDipole}) 
plays the role of the deformation. For small deformation where
$e^{\rho_0}\ll 1$,
the corresponding subsectors have the following deviation from the LST
\bea 
(E-J)_{\rm NC\;LST}&=&(E-J)_{\rm LST}+
{e^{2\rho_0}\over 2}\sum_n\left[{N|n|\over J}N_{n}^{(6)}+
{N^2n^2\over J^2}\left({N_n^+\over \omega^+_n}+{N_n^-\over \omega^-_n}\right)
\right],\cr &&\cr
(E-J)_{\rm DI\;LST}&=&(E-J)_{\rm LST}+
{e^{2\rho_0}\over 2}\sum_n\left[(E-J)_{\rm LST}+
{N^2n^2\over J^2}\left(N_n^++N_n^-\right)\right],
\eea
up to second order in fluctuations and first order in the 
noncommutative/dipole deformations.

The next step, of course, would be to write the 
explicit form of the corresponding operators. These operators must be made
out of the field content of LST theory. For example consider type IIA 
NS5-brane. In this case the bosonic part of the theory contains an
anti-symmetric
self-dual two form $B_{\alpha\beta},\alpha=0,\cdots,5$, 2 complex scalars
and one real scalar (dilaton). Suppose $Z_1$ is a complex scalar carrying
one unit of $U(1)$ charge. Therefore one would expect that the vacuum 
to be constructed from ${\rm Tr}(Z_1^J)$. Of course, since the theory
is not conformal one can not identify the energy with dimension of 
the operator. Nevertheless, following \cite{Berenstein:2002jq} one 
could guess that the 
other stringy states can be constructed by inserting either
the other scalar or $B_{\alpha\beta}$. It would be quite interesting and
also important to study these operators in detail. Among all other things
this could increase our knowledge about LST theory.
We hope to address this question in future.

\vspace*{.4cm}

{ \bf Acknowledgements}

We would like to thank S. Parvizi  for useful discussions. We would also
like to thank Carlos Nunez and Mohammad M. Sheikh-Jabbari for useful comments.

\end{document}